\documentclass[conference]{IEEEtran}
\IEEEoverridecommandlockouts

\usepackage{cite}
\usepackage{amsmath,amssymb,amsfonts}
\usepackage[ruled,vlined,linesnumbered]{algorithm2e}
\usepackage{algorithmic}

\usepackage{graphicx}
\usepackage{textcomp}
\usepackage{xcolor}
\usepackage{booktabs}
\usepackage{amssymb}
\usepackage{pifont}
\usepackage{fancyhdr}

\def\BibTeX{{\rm B\kern-.05em{\sc i\kern-.025em b}\kern-.08em
    T\kern-.1667em\lower.7ex\hbox{E}\kern-.125emX}}

\fancypagestyle{firstpage}{%
 \fancyhf{}%
 \fancyhead[L]{CPSAT 2024 - The 5th CPSSI International Symposium on Cyber-Physical Systems (Applications and Theory)}
  
 \fancyfoot[L]{%
  \normalsize{XXX-X-XXXX-XXXX-X/XX/\$XX.00 ~\copyright2024 IEEE}
 }%
}

\begin{document}

\title{\huge{LIMO: Load-balanced Offloading with MAPE and Particle Swarm Optimization in Mobile Fog Networks}}

\author{

 \IEEEauthorblockN
 {
 Yasaman Seraj\textsuperscript{1,2},
 Soheil Fadaei\textsuperscript{1,2},
 Bardia Safaei\textsuperscript{1,2},
 Ali Javadi\textsuperscript{3},\\
 Amir Mahdi Hosseini Monazzah\textsuperscript{3},
 Ali Mohammad Afshin Hemmatyar\textsuperscript{2}
 }
    \IEEEauthorblockA{
    \textsuperscript{1}Reliable And Durable IoT Applications \& Networks (RADIAN) Research Laboratory,\\
    \textsuperscript{2}Department of Computer Engineering, Sharif University of Technology, Tehran, Iran\\
    \textsuperscript{3}Department of Computer Engineering, Iran University of Science and Technology (IUST), Tehran, Iran
    \vspace{-10pt}}
}



\maketitle
\thispagestyle{firstpage}
\begin{abstract}
Fog computing is essentially the expansion of cloud computing towards the network edge, reducing user access time to computing resources and services. Various advantages attribute to fog computing, including reduced latency, and improved user experience. However, user mobility may limit the benefits of fog computing. The displacement of users from one location to another, may increase their distance from a fog server, leading into latency amplification. This would also increase the probability of over utilization of fog servers which are located in popular destinations of mobile edge devices. This creates an unbalanced network of fog devices failing to provide lower makespan and fewer cloud accesses. One solution to maintain latency within an acceptable range is the migration of fog tasks and preserve the distance between the edge devices and the available resources. Although some studies have focused on fog task migration, none of them have considered load balancing in fog nodes. Accordingly, this paper introduces LIMO; an allocation and migration strategy for establishing load balancing in fog networks based on the control loop MAPE (Monitor-Analyze-Plan-Execute) and the Particle Swarm Optimization (PSO) algorithm. The periodical migration of tasks for load balancing aims to enhance the system's efficiency. The performance of LIMO has been modeled and evaluated using the Mobfogsim toolkit. The results show that this technique outperforms the state-of-the-art in terms of network resource utilization with 10\% improvement. Furthermore, LIMO reduces the task migration to cloud by more than 15\%, while it reduces the request response time by 18\%.
\end{abstract}


\section{Introduction}

The Internet of Things (IoT) represents a paradigm shift in the way devices interact and communicate with each other and the external environment. IoT enables the integration of various sensors, actuators, and data analytics tools to create a cohesive network that provides real-time data and insights \cite{safaei2022introduction}. This technology offers numerous advantages, including enhanced automation, improved efficiency, and better decision-making capabilities. For instance, it plays a pivotal role in smart cities, healthcare, agriculture, and industrial automation, where it facilitates remote monitoring, predictive maintenance, and resource management \cite{gubbi2013internet,shirbeigi2021cluster}.

Despite its advantages, the vast amount of data generated by IoT devices can lead to challenges related to data storage, processing, and management. Moreover, the need for low-latency responses can be difficult to achieve with traditional centralized computing models. These challenges necessitate the reliance on cloud computing, which provides the necessary computational power and storage capabilities to handle large-scale data processing and analytics \cite{dastjerdi2016fog}. However, the reliance on cloud computing comes with its own set of drawbacks, including latency issues, bandwidth constraints, and potential privacy concerns. As IoT applications continue to proliferate, the sheer volume of data being produced and the need for immediate processing capabilities often render edge nodes insufficient for handling such tasks. Consequently, there is a growing need to leverage other cloud-based technologies to meet these demands \cite{yi2015survey}.

To address the limitations of both edge and cloud computing, fog computing has emerged as a viable solution. Fog computing extends cloud services to the edge of the network, providing a decentralized computing infrastructure that brings data storage and processing closer to the source of data generation. By processing data closer to the source, fog computing reduces the need to transfer large volumes of data to centralized cloud servers, thus alleviating bandwidth congestion and minimizing latency. Fog computing is characterized by its support for mobility, wide geographical distribution, and the ability to deliver low-latency interactions. Meanwhile, ensuring seamless service amidst user mobility requires efficient task migration across nodes. Mobility and limited IoT node coverage necessitate multi-hop support in fog infrastructures, directly impacting Quality of Service (QoS) \cite{chiang2016fog}. Therefore, dynamic service migration is vital to bring the services closer to mobile users. Nevertheless, mobility-aware migration faces challenges such as distance between end node and fog devices, service size, bandwidth, and fog node utilization \cite{rejiba2019a}.

Imposing Heavy loads on fog nodes can lead to several critical issues, including increased latency, higher energy consumption, and potential system failures. Overloaded fog nodes struggle to process tasks efficiently, resulting in delays that can degrade the quality of service (QoS) for time-sensitive IoT applications. Additionally, the excessive computational demands can cause thermal throttling and increased power usage, further exacerbating the strain on the system \cite{chiang2016fog}. These problems highlight the necessity for effective load-balancing mechanisms, especially in mobile applications, where the resource allocation for task execution turns into complicated decision-making. Load balancing can mitigate these issues by distributing tasks evenly across available fog nodes, preventing any single node from becoming a bottleneck. This approach enhances the overall system performance by reducing latency, optimizing energy consumption, and increasing the reliability and resilience of the network \cite{stojmenovic2014fog}. By ensuring that no single fog node is overwhelmed, load balancing helps maintain a stable and efficient computational environment, crucial for the success of IoT applications.

To tackle the problem of imbalanced utilization of fog devices in mobile applications, which leads to a higher offloading rate to the cloud, and increased Task Completion Times (TTC), this paper introduces \textbf{LIMO}; an allocation and task migration strategy for fog computing frameworks, based on the MAPE control loop along with the Particle Swarm Optimization (PSO). PSO excels in load balancing for task offloading in mobile fog networks due to its ability to handle non-linear, high-dimensional problems and avoid local optima. Additionally, PSO's flexibility and ease of implementation make it a versatile and efficient choice. The main goal of LIMO is to reduce the makespan of tasks by evenly balancing the load and utilization among fog devices. These goals have been achieved by selecting the optimal node among the available nodes, taking into account the constraints of the problem, including the number of available fog nodes, their resources including CPU, memory, bandwidth, the distance of users from the cloud, and the mobility characteristics of the end users. A metaheuristic approach will be employed for node selection. The performance of LIMO is investigated and compared with the state-of-the-art in terms of various metrics in the Mobfogsim simulation environment \cite{puliafito2020a}. The results demonstrate that LIMO significantly improves network resource utilization by 10\%, reduces task migration to the cloud by more than 15\%, and decreases request response time by 18\% compared to the state-of-the-art.

In the continuation of this paper, Section II explains edge, fog, and cloud computing concepts. The literature will be reviewed in section III. Section IV will provide a detailed explanation of the proposed method. The experimental results will be discussed in Section V, and the paper will be concluded in Section VI.

\section{Preliminaries}
Cloud computing, fog computing, and edge computing are pivotal technologies in modern computing architectures, each playing a unique role in the processing and management of data. Meanwhile, cloud computing is a model that provides ubiquitous, convenient, on-demand network access to a shared pool of configurable computing resources, such as networks, servers, storage, applications, and services \cite{mell2011nist}. This paradigm enables users to leverage vast computational power and storage without investing in physical infrastructure, thus offering scalability, flexibility, and cost efficiency. Major cloud service models include Infrastructure as a Service (IaaS), Platform as a Service (PaaS), and Software as a Service (SaaS), each catering to different levels of user requirements.

Fog computing extends cloud computing capabilities to the edge of the network, closer to the data sources and end-users \cite{bonomi2012fog}. This intermediate layer between cloud data centers and edge devices aims to reduce latency, enhance data security, and improve real-time data processing capabilities. By processing data locally or near the data source, fog computing mitigates the bandwidth constraints and delays associated with cloud computing. This paradigm is particularly beneficial for applications requiring immediate processing and low-latency responses, such as autonomous vehicles, smart grids, and industrial IoT systems.

On the other hand, edge computing involves processing data at or near the data source, reducing the need to transmit vast amounts of data to centralized cloud servers \cite{shi2016edge}. This approach minimizes latency and bandwidth usage, providing faster data processing and response times. Edge computing is crucial for applications where real-time decision-making is critical, such as augmented reality, telemedicine, and real-time analytics. By enabling data processing at the edge, this paradigm complements fog computing by handling time-sensitive tasks and reducing the load on fog and cloud infrastructures.

The interconnection between cloud, fog, and edge computing forms a comprehensive hierarchical architecture. Edge computing handles immediate, latency-sensitive tasks directly at the data source. Fog computing serves as an intermediary, processing data that require quick responses but are less time-critical than those handled by edge computing. The cloud provides substantial computational power and storage for tasks that can tolerate higher latency and do not require immediate processing. This hierarchical approach ensures that data is processed at the most appropriate level, optimizing resource utilization and enhancing overall system performance. It worths mentioning that edge computing is often erroneously called fog computing. Although these paradigms move the computation and storage to the edge of the network, they are not identical. The major differences of these two technologies are indicated in Table. \ref{table:compare}.
 \begin{table}[t]
    \centering
        \caption{A comparison between fog and edge computing paradigms.}
    \includegraphics[scale=0.53]{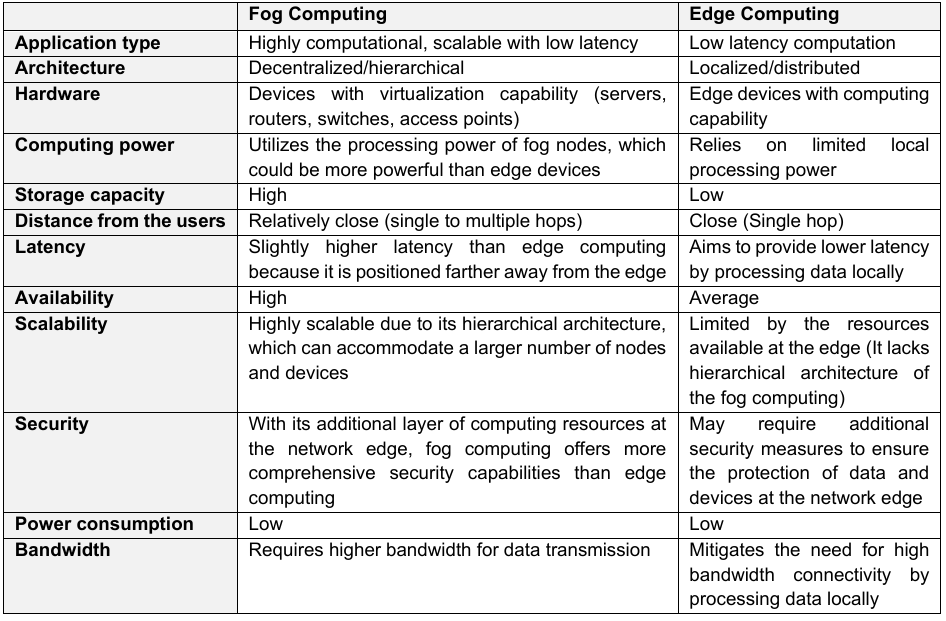}
    \label{table:compare}
    \vspace{-18pt}
\end{table}

In a general cloud/fog infrastructure, offloading refers to the transfer of computational tasks from resource-constrained devices residing in the lower levels of the architecture (IoT edge) to more powerful fog or cloud nodes. This process is crucial for managing the computational load, reducing latency, and enhancing energy efficiency in resource-constrained environments \cite{satyanarayanan2017emergence}. Offloading is especially important in mobile fog networks where user mobility can lead to varying load conditions across fog nodes. Efficient offloading strategies ensure balanced load distribution, preventing node overload and underload, and maintaining optimal system performance.
 
\section{Related Studies}
There are four major families of algorithms related to this study. These include approximate, exact, fundamental, and hybrid algorithms.

\textbf{Approximate Algorithms}: There are multiple types of algorithms in this context, which are stochastic, probabilistic,
statistic, heuristic and meta-heuristic \cite{kashani2022a}. Heuristic algorithms are designed based on "experience" for specific optimization problems, aiming to find the best solution through "trial and error" in an optimal timeframe. Solutions derived from heuristic approaches may not always be the best or optimal, but they often surpass mere guesses. These approaches leverage the characteristics of the problem at hand. Given that exact approaches require significant time to achieve optimal solutions, heuristic approaches are preferred for obtaining near-optimal solutions within a reasonable timeframe. Some heuristic methods explored in research include hill climbing, minimal conflicts, and the analytic hierarchy process\cite{sofla2022a}. A mobility-aware autonomic approach for the migration of application modules in fog computing environments (MAMF) has been proposed in \cite{martin2020a}. This hybrid solution employs the MAPE loop concepts alongside a genetic algorithm to manage carrier migration within the fog environment, ensuring the timely distribution of requests and maintaining QoS for the end-users. In \cite{baburao2021a} the authors proposed an Enhanced Dynamic Resource Allocation Method (EDRAM) based on the Particle Swarm Optimization (PSO), which improves the Quality of Experience (QoE) and reduces delay. In \cite{yang2020a}, to further optimize processing delay in the network architecture, the authors have introduced a load-balancing strategy within a fog computing network. Given the robust global search capabilities of the Bat Algorithm (BA), this study employs the BA algorithm to address the optimization problem in a medical big data scenario. In another study, the FGWHO algorithm, based on the whale optimization algorithm, is proposed for a wireless sensor network connected to a micro-grid and operating in a fog computing environment \cite{karthik2021a}. This approach focuses on enhancing the network's lifespan by optimizing the routing of a system connected to the network, thereby preventing malfunction. In \cite{qun2021a}, an energy-aware method for load balancing in fog-based Vehicular Ad-hoc Networks (VANETs) is presented, utilizing a hybrid optimization algorithm that combines Ant Colony Optimization (ACO) and Artificial Bee Colony (ABC) algorithms. Authors in \cite{singh2020a} introduce a fuzzy logic-based load-balancer utilizing various levels of fuzzy control settings and design within fog networks. The fuzzy logic model is employed for link analysis as interconnections to manage traffic. According to the results, a fuzzy logic controller can handle many complex or overlapping scenarios that may arise in real-life situations. In study \cite{beraldi2020a}, a fully distributed algorithm for load balancing based on random exploration of neighbor states is examined. This study considers the impact of delay during the exploration phase and analyzes the effect of outdated load information. The results indicate that making scheduling decisions based on state information, even with slight delays compared to service time, significantly reduces the effectiveness of load balancing. Authors in \cite{motamedhashemi2023data} proposed a throughput and deadline-aware task scheduling mechanism for time-sensitive fog frameworks, utilizing a genetic algorithm. The researchers employed genetic optimization by encoding potential solutions into chromosomes. By applying gene mutation and two-point crossover techniques, the proposed method achieves a high guarantee ratio while maintaining a low makespan.
In \cite{siyadatzadeh2023relief}, a task assignment strategy based on machine learning is proposed. This approach utilizes reinforcement learning to identify suitable nodes for executing primary and backup tasks. The method demonstrates strong performance in dynamic environments by balancing communication delay and workload across each fog device.

\textbf{Exact Algorithms:}
In a study conducted in \cite{cui-a}, an algorithm named Minimum Response Time Balancing (MRTB) is proposed. In this algorithm, the Dijkstra algorithm is employed to find the shortest path with a single source and without negative weights. The authors suggest adapting the Dijkstra algorithm to address load balancing in fog computing and vehicular environments.
A minimized bandwidth cost and efficient resource management in a collaborative fog-cloud computing environment has been investigated in \cite{maswood2020a}. In this study, an optimization model based on mixed-integer linear programming is proposed to minimize the combined objective function.

\textbf{Fundamental Algorithms:}
In \cite{alqahtani2021a}, a reliable scheduling approach for allocating customer requests to resources in fog-cloud environments is introduced. This approach, named Load Balanced Service Scheduling Approach (LBSSA), considers load balancing among resources by categorizing requests as urgent, important, and tolerant to latency during their allocation to resources. Additionally, request scheduling in the proposed approach takes into account resource failure rates to ensure high reliability for requested services. Authors in \cite{asghar2021a} present a fog-based health monitoring system called LBS, designed to minimize latency and optimize network usage. Additionally, a novel load balancing scheme is proposed to distribute the load among fog nodes effectively, particularly when the health monitoring system is deployed on a large scale.

\textbf{Hybrid Algorithms:}
To achieve load balancing in fog networks, hybrid algorithms combining various approaches such as heuristic, exact, and foundational algorithms may be also employed. In research presented in \cite{talaat2020a}, a load balancing and optimization strategy (LBOS) is proposed. This approach employs a dynamic resource allocation method based on reinforcement learning and genetic algorithms. LBOS actively monitors network traffic, gathers load information from each service provider, manages incoming requests, and distributes them evenly among available service providers using the dynamic resource allocation technique.
In \cite{yan-a}, a trade-off has been achieved between latency and energy consumption for fog nodes. Furthermore, the problem of minimizing maximum load in homogeneous fog networks is formulated, demonstrating it to be an NP-hard problem. Subsequently, a greedy algorithm named YA is proposed to address the critical task offloading problem in fog nodes, assigning tasks to fog nodes with minimal load across the network. Additionally, to tackle the selfishness consideration of fog nodes, a coalition-based algorithm is suggested to encourage lightly loaded fog nodes to share their resources in order to reduce maximum load.

\section{Detailed Description of LIMO}
\begin{figure}[t]
\centerline{\includegraphics[scale=.35]{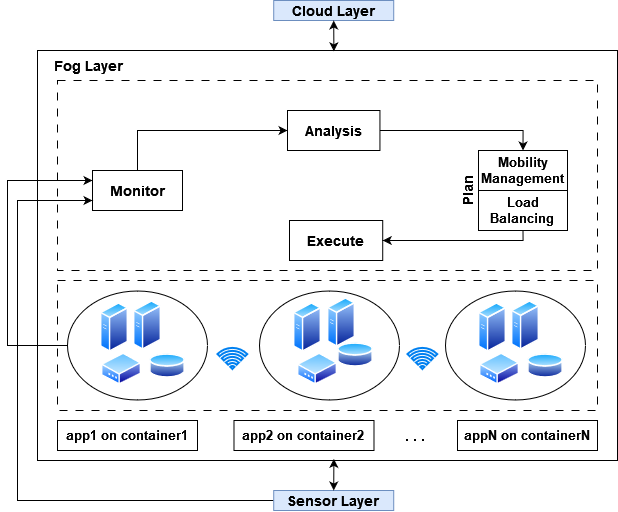}}
\caption{Architecture of LIMO based on the MAPE control loop.}
\label{4-1-mage}
\vspace{-10pt}

\end{figure}

The primary objective of this method is to develop a framework for the automatic transfer of user devices. Within this framework, devices operate autonomously by leveraging decisions made through the MAPE control loop. Fig \ref{4-1-mage} illustrates the architecture of this method. 
The MAPE control loop facilitates mobility support through efficient coordination of application modules. Implemented in the fog layer, the MAPE loop comprises four stages: Monitoring, Analysis, Planning, and Execution. Initially, the system assesses the necessary number of fog nodes. For each incoming user request,the MAPE control loop is executed at specified time intervals to determine whether the migration of application modules is required.

\begin{itemize}
    \item \textbf{Monitoring:} In this phase, information is gathered through monitoring tools. The user monitor tracks the location history of requesting users, while the resource monitor oversees the status of resources. This collected information is stored in the knowledge base for subsequent use.
   \item \textbf{Analysis:} This phase processes the data collected during the monitoring phase. The analysis involves predicting the user's next location, which is then forwarded to the planning phase.\item \textbf{Planning:} Based on the analysis, the system plans the migration. If migration is required, the planning phase identifies the optimal node for relocation.The data analysis determines whether migration is necessary. If migration is deemed necessary, the appropriate node is selected using the Particle Swarm Optimization (PSO) algorithm.\item \textbf{Execution:} In this phase, the planned migration is carried out. The execution involves the actual transfer of application modules to the selected node.
\end{itemize}
            

\begin{algorithm}[t]
\footnotesize
\SetAlgoLined
 \KwIn{$X_1, \ldots X_t $ and resource utilization values}
 \KwOut{Monitoring the time-based location history of the user for being used in Analyze phase}
 \textbf{Begin}\\
Monitor $\{ X_1, X_2, \ldots, X_t \}$\\
Monitor $\{$Resource utilization values$\}$\\
\textbf{End}
\caption{Monitoring Phase}
\end{algorithm}


        
    

\begin{algorithm}[t]
\footnotesize
\SetAlgoLined
\KwIn {Time-based location history of a particular mobile device until time $t$}
\KwOut{ Forecast location of the mobile device in $t+1$}
  \textbf{Begin}\\
{Read the values of $(x_1,y_1), (x_2,y_2), \ldots (x_t,y_t)$}\\
\textbf{Return}: {Forecast value of $(x_{t+1}, y_{t+1})$ }\\
\textbf{End}
\caption{Analyze Phase}
\end{algorithm}

\begin{algorithm}[t]
\footnotesize
\SetAlgoLined
\KwIn {Location at time $t+1$ : $(x_{t+1}, y_{t+1})$}
\KwOut{ decision to migrate or not}
\textbf{Define:} {$D$ = Distance between $FN_{current}$ and user}\\
\If{ $600 < D < 1000$ }{
    \If{$Candidate(FN) \neq NULL$}{$FN_{Next} = Candidate(FN)$}
    \Else {$FN_{Next} = Cloud$}
}
\caption{First Part of the Planning: Mobility Management}
\end{algorithm}

\begin{algorithm}[t]
\footnotesize
\SetAlgoLined
\KwIn {Set of \{$FN_1, FN_2 , \ldots, FN_n$\} and Resources of each $FN$ and taken resources of each task assigned to $FN$}
\While{TRUE}{
\textbf{Do:} {Calculate the total time for tasks to be completed by $FNs$}\\
\textbf{Do:} {Calculate the utilization of each Fog Node during the time of performing assigned tasks}\\
\textbf{Do:} {Calculate the average utilization of Fog Nodes}\\
\textbf{Do:} {Calculate the standard deviation ($SD_n$) of load of $FN$s}\\
\If{$SD_n > threshold$}{Unbalanced}
\Else{Almost Balanced, Break;}
\If{Unbalanced}{Execute Phase: Move tasks from $FN$ with more load to $FN$ with less load}
}
\caption{Second Part of the Planning: Load Balancing}
\end{algorithm}

\begin{algorithm}[t]
\footnotesize
\SetAlgoLined
\KwIn {Migration destination $FN$ or $cloud$}
\KwOut{Migrate tasks}
\textbf{Do:} {Migrate Task $N$ to another $FN$ of $cloud$ that has been determined by planning phase. }\\
\textbf{End}

\caption{Execution Phase}
\end{algorithm}

Algorithms 1 to 5 show the procedure that is done during the MAPE control loop in our method. By employing the MAPE control loop, the proposed method ensures efficient and autonomous management of user device transfers within the fog computing environment\cite{martin2020a}.

Load balancing is a critical aspect of scheduling migratory modules in a fog environment. It involves distributing the workload across multiple servers to optimize resource utilization, thereby achieving the best possible response time and throughput in the entire network. Therefore, an effective load-balancing mechanism can harmonize the major objectives in a fog infrastructure amidst the existing tradeoffs. Meanwhile, the resource utilization of a fog node is defined as the ratio of the node's active time to the total completion time of the migratory modules it handles. Generally speaking, by employing a mechanism aimed at reducing the overall completion time of migratory modules, it is expected to observe higher resource utilization. However, the increase in the utilization must be controlled to avoid reaching 1. Hence, applying an appropriate load balancing mechanism seems to be a challenging optimization problem. To formulate this optimization problem, consider a set of migratory modules denoted as $\{T_1, T_2, T_3, ..., T_n\}$ and a set of fog nodes represented as $\{FN_1, FN_2, ..., FN_m\}$. Let the completion time of fog node $FN_j$ be denoted as $CT_j$. Equations \ref{4-4} and \ref{4-5} define the overall makespan and the utilization of fog nodes, respectively. The objective is to minimize the overall makespan while maximizing the resource utilization, as specified in Equation \ref{4-5}. This enables us to get the most out of the deployed fog devices in the network to minimize the makespan. The average utilization is defined by Equation \ref{4-6}.
\begin{equation}
{Makespan} = \max\{{CT}_j \mid j = 1, 2, \ldots, m \}
\label{4-4}
\end{equation}
\begin{equation}
{Util}_{{FN}_j} = \frac{CT_j}{{Makespan}}
\label{4-5}
\end{equation}
\begin{equation}
{Util_Avg} = \frac{\sum_{j=1}^{m} {Util}_{{FN}_j}}{m}
\label{4-6}
\end{equation}
The $TTC_i$ metric is calculated as the sum of the time required to transfer the application module $app_i$ to the destination fog node $FN_j$ and the time required to process the request by the application module at the destination $FN_j$. $TTC_i$ for a migratory module consists of two components. The first component is the time necessary to transfer the application module ($MT_{app_{ij}}$) from the current fog node $FN_k$ to the designated destination fog node $FN_j$, with $D_{pd}$ representing the propagation delay. Migration time represents the time spent transferring the relevant memory contents over the available network bandwidth. This time is calculated as per the given equation. The second component, $PT_ij$, evaluates the time spent processing the request sent to the application module at the destination fog node. In other words, $PT_ij$ is the time required for module '$i$' to process or service the request at fog node '$j$'. Equation \ref{4-7} shows how $TTC$ is calculated.
\begin{equation}
{TTC}_i = ({MT}_{{app}_{ij}} + D_{pd}) + {PT}_{ij}
\label{4-7}
\end{equation}
where \({MT}_{{app}_{ij}}\) is the migration time for application \({app}_i\) given by:
\begin{equation}
{MT}_{{app}_{ij}} = \frac{\text{required memory for}\; {app}_i}{\text{bandwidth of}\; {FN}_j }
\label{4-7-1}
\end{equation}

Considering the constraints of Equations \ref{4-8} to \ref{4-10}, the fitness function of the PSO algorithm is defined by Equation \ref{4-11}. According to this equation, a node has a better position if it has a higher fitness value. The weights $w_1$ and $w_2$ vary based on the load balancing.The characteristics of a fog node are represented as a tuple "$<fn(cpu)_{j}^{cap}, fn(mem)_{j}^{cap}, fn(bw)_{j}^{cap}>$", where $fn(cpu)_{j}^{cap} $denotes the CPU capacity, $fn(mem)_{j}^{cap}$ indicates the memory capacity, and $fn(bw)_{j}^{cap}$ represents the available bandwidth of the $j-th$ fog node. Similarly, migrated tasks are represented as a tuple "$<app(cpu)_{i}^{cap}, app(mem)_{i}^{cap}, app(bw)_{i}^{cap}>$", where $app(cpu)_{i}^{cap}$ denotes the CPU requirement, $app(mem)_{i}^{cap}$ indicates the memory requirement, and $app(bw)_{i}^{cap}$ represents the bandwidth requirement of the user's task.

\begin{equation}
\forall {FN}_j \quad \sum_{i=0}^{n} {app(cpu)}_i^{{req}} \leq {FN(cpu)}_j^{{cap}} \quad 
\label{4-8}
\end{equation}
\begin{equation}
\forall {FN}_j \quad \sum_{i=0}^{n} {app(mem)}_i^{{req}} \leq {FN(mem)}_j^{{cap}} \quad 
\label{4-9}
\end{equation}
\begin{equation}
\forall {FN}_j \quad \sum_{i=0}^{n} {app(bw)}_i^{{req}} \leq {FN(bw)}_j^{{cap}} \quad 
\label{4-10}
\end{equation}
\begin{equation}
\text{Fitness function} = w_2 \cdot {Avg Util}_{{norm}} - w_1 \cdot {TTC}_{{inorm}} \quad 
\label{4-11}
\end{equation}

Normalizing the values in Equation \ref{4-11} is achieved using Equations \ref{4-12} and \ref{4-13}. The min/max algorithm is employed for normalization to ensure a uniform data distribution.
\begin{equation}
\label{4-12}
TTC_{i_{norm}} = \frac{TTC_i - TTC_{min}}{TTC_{max} - TTC_{min}} \quad 
\end{equation}

\begin{equation}
AvgUtil_{norm} = \frac{AvgUtil_i - AvgUtil_{min}}{AvgUtil_{max} - AvgUtil_{min}} \quad 
\label{4-13}
\end{equation}

By selecting an appropriate objective function, solutions with lower makespan and higher resource utilization can be chosen. To offload tasks from an overloaded fog node, a set of surrounding fog nodes should be considered for load balancing. Load imbalance in the fog environment can be calculated by computing the standard deviation of the system load, as given by Equation \ref{4-14}.
\begin{equation}
\label{4-14}
\sigma = \sqrt{\frac{1}{m} \sum_{i=1}^m (LFN_i - AVGLFN)^2} \quad 
\end{equation}
In this equation, \( m \) is the number of fog nodes in a group, \( LFN_i \) is the processing load of the fog nodes, and \( AVGLFN \) is the average processing load of the fog nodes in a group.
By comparing the standard deviation with the threshold, the load balance status can be categorized into three states:
\begin{itemize}
\item  Almost balanced (\, 0 $<$ \text{standard deviation} $<$ \text{threshold} \,)
\item Unbalanced (\, \text{standard deviation} $\ge$ \text{threshold} \,)
\item  Balanced (\, \text{standard deviation} = 0 \,)
\end{itemize}
If the system is balanced or almost balanced, \( w_1 = 1 \) and \( w_2 = 0 \). If the system is unbalanced, \( w_1 = 0.5 \) and \( w_2 = 0.5 \).
The detailed steps of the algorithm are as follows:
\begin{enumerate}
  
\item Defining particles, position vectors, and velocity vectors.
For a problem with \( n \) tasks, each particle is defined as an \( N \)-dimensional vector \( X = (X_1, X_2, ..., X_n) \) where \( X_i \) \((i \in \{1, 2, ..., n\})\) represents the fog node where the \( i \)-th task is processed. The positions and velocities of the particles are initialized randomly.

\item  Calculating the fitness function and determining \( pbest \). The fitness value of each particle is calculated using Equation \ref{4-11}. By comparing the current fitness value of each particle with its \( pbest \) value, the best position of the particle (\( pbest \)) is determined. If the current fitness value is greater than the current \( pbest \), then the \( pbest \) is updated to the current fitness value.

\item  Determining \( gbest \)
By comparing the current fitness values of the entire population, the highest fitness value is identified as the global best position (\( gbest \)).

\item  Updating the position and velocity vectors of the particles.  
The velocity vector of each particle is updated according to Table \ref{table:4-1}. The position vector of each particle is updated using Equation \ref{2-2}.
\begin{equation}
\label{2-2}
 x_{i}^{(k+1)}= x_{i}^{k}+ V_{i}^{(k+1)}   
\end{equation}
\item  Termination condition. 
Steps 2 to 4 are repeated until the maximum number of iterations is reached.
\end{enumerate}
This algorithm is simulated with the parameters listed in Table \ref{table:4-1}. The values of \( c \) are chosen based on similar studies.


\begin{table}[t]
\caption{Parameters used in the PSO algorithm}
\centering
 \includegraphics[scale=0.23]{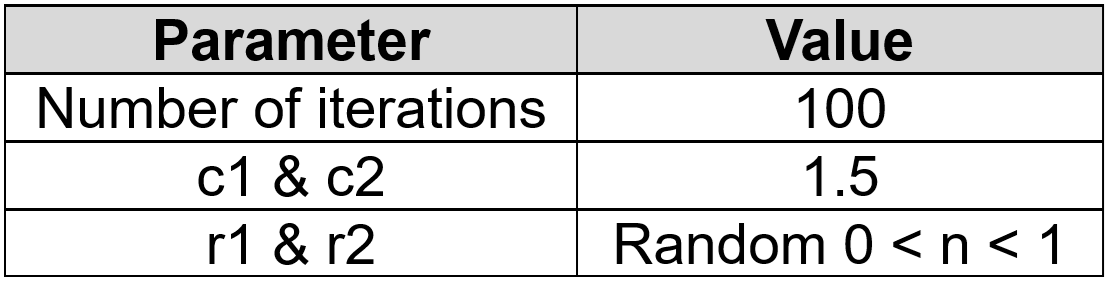}
    \label{table:4-1}
    \vspace{-18pt}
\end{table}

\section{System Setup and Experimental Evaluations}

Experiments were designed to evaluate and analyze the performance of the proposed model in various scenarios. These experiments utilized a fog simulation toolkit called MobFogSim to simulate the fog environment and cloud resources, as well as a real-time vehicular mobility dataset to emulate the real-time movement of users.
\begin{table}[t]
\centering
\caption{Simulation Parameters}
\label{tab:sim_details2}
\includegraphics[scale=0.23]{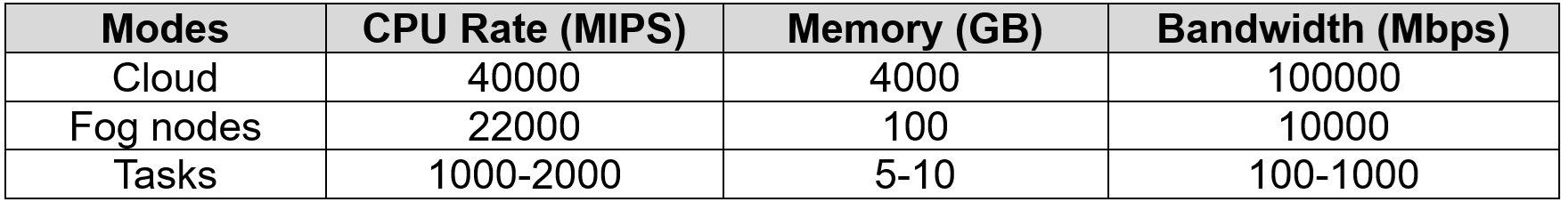}
    \vspace{-15pt}
\end{table}

MobFogSim, a simulator based on iFogSim, enables the modeling of node mobility and service migration. In this simulator, the impact of mobility on service migration is examined using an experimental approach. MobFogSim can consider user mobility, wireless connectivity, and the migration process. It is capable of employing a user-defined migration strategy. MobFogSim provides a useful basis for supporting fog computing in applications where users are mobile and where a migration strategy for transferring state or data in the clouds is needed. Although many simulators exist to evaluate the behavior and performance of applications in a fog computing environment, none allow for the evaluation of service migration solutions to support mobility. MobFogSim overcomes this limitation \cite{puliafito2020a}.

The performance of the proposed method was evaluated in a fog computing environment with varying numbers of application modules and fog nodes. The details of the simulation parameters are provided in Table \ref{tab:sim_details2}.

To validate the efficiency of the proposed method, a real-time mobility dataset was used \cite{l-a}. This dataset was collected by the General Departmental Council of Val de Marne, located in Creteil, France. The General Departmental Council is a regional agency responsible for controlling and coordinating the transportation system in France. The dataset includes the tracking of vehicle flows in the city during two peak periods: two hours in the morning and two hours in the evening, and it considers various types of vehicles in the experiments.
The synthetic tracking scenario includes a roundabout with 6 entry/exit points, 2 or 3-lane roads, 1 bus lane, 4 lane-changing points, and 15 traffic lights. This dataset contains approximately 10,000 trips during the busy two-hour periods in the morning (7 AM to 9 AM) and evening (5 PM to 7 PM).
The dataset features tracking attributes at each time interval (with a granularity of 1 second), including vehicle lane, vehicle angle, vehicle type (vehicle or bus), vehicle position, vehicle coordinates in the 2D plane (x and y coordinates in meters), vehicle speed (in meters per second), and vehicle ID. The dataset is provided in two compressed CSV files (one for the morning and one for the evening) and can be downloaded from the website \cite{unknown-a}. The morning dataset includes 857,136 positions (samples), while the evening dataset includes 902,735 positions.
When a user moves out of the coverage area of a fog node, to provide effective support for user mobility, application modules must be transferred from the current fog node to another fog node that is closer to the user's current location. This ensures better coverage for the user as they move. Selecting an appropriate target fog node for the application module when multiple fog nodes are available is challenging.
In this study, the MAPE control loop continuously runs, monitors conditions and performs migrations as necessary.
\begin{figure}[t]
\centerline{\includegraphics[scale=.16]{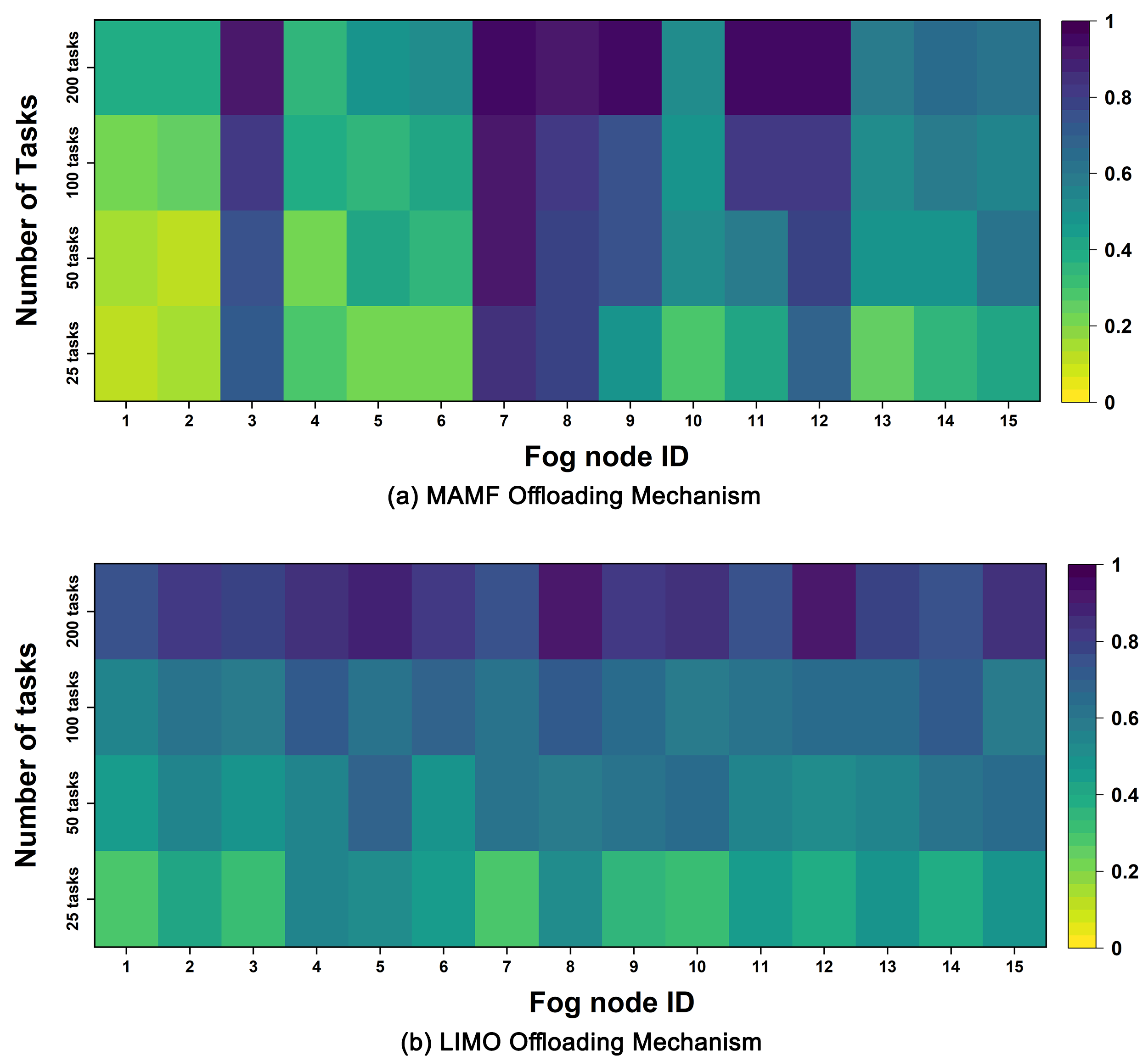}}
\caption{Utilization of 15 Fog nodes with different number of Tasks. }
\label{4_heatmap}
\vspace{-10pt}
\end{figure}
\begin{figure}[t]
\centerline{\includegraphics[scale=.30]{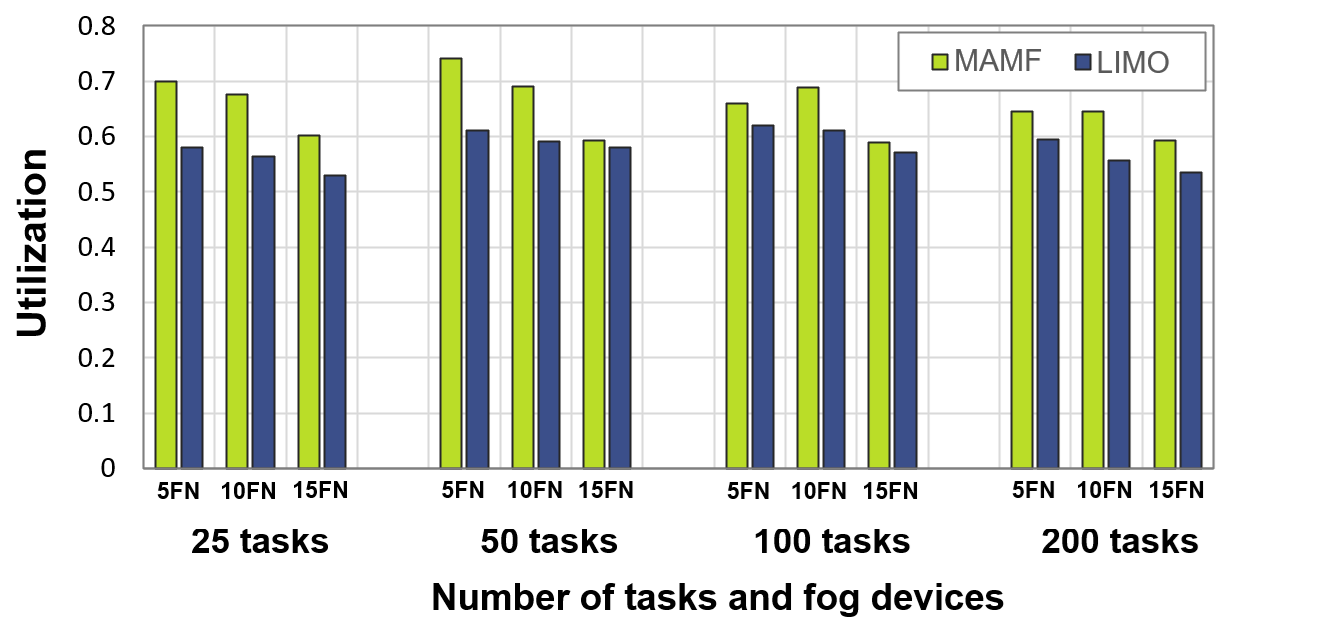}}
\caption{Average utilization of fog nodes in various fog scenarios.}
\label{1_util}
\vspace{-16pt}
\end{figure}

The performance of LIMO has been compared and evaluated against the MAMF offloading mechanism. Fig. \ref{4_heatmap} presents a heatmap illustrating the utilization of each fog node for task counts of 25, 50, 100, and 200. In this heatmap, darker cells indicate higher utilization and greater fog node activity, while green and yellow cells represent lower utilization and less fog node activity. By analyzing the heatmap, it is evident that LIMO achieves better utilization distribution. The nodes are more uniformly and progressively loaded, demonstrating improved load balancing. According to Fig. \ref{4_heatmap}(a), by using MAMF, some of the nodes , e.g, nodes 3, and 7 are overloaded due to their high-popularity location among the mobile edge devices, which has made them a popular target for task offloading while the other nodes are underloaded. However, as indicated in Fig. \ref{4_heatmap}(b), due to using PSO, LIMO has overcome this challenge and provided evenly balanced loads. The main reason for this behavior is the lack of consideration of load balancing in MAMF. As a consequence of this overloading, a substantial number of tasks will be migrated to the cloud, leading to a higher average makespan and elevated network resource consumption.

 \begin{figure}[t]
\centerline{\includegraphics[scale=.32]{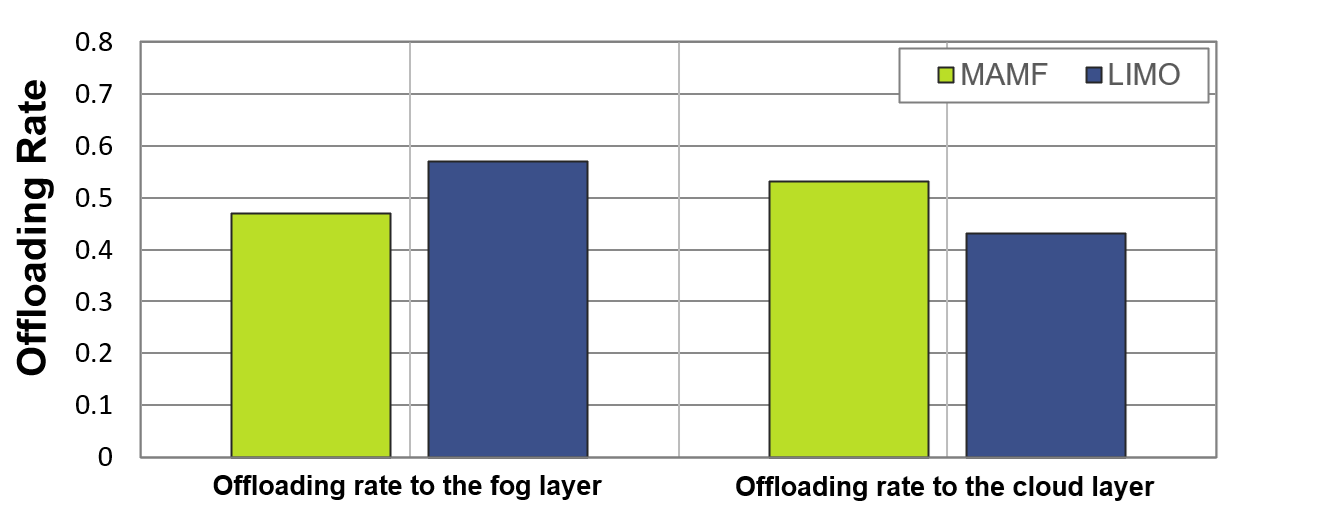}}
\caption{Average Offloading rate of the tasks.}
\label{2_OL}
\vspace{-16pt}
\end{figure}

\begin{figure}[b]
\centerline{\includegraphics[scale=.30]{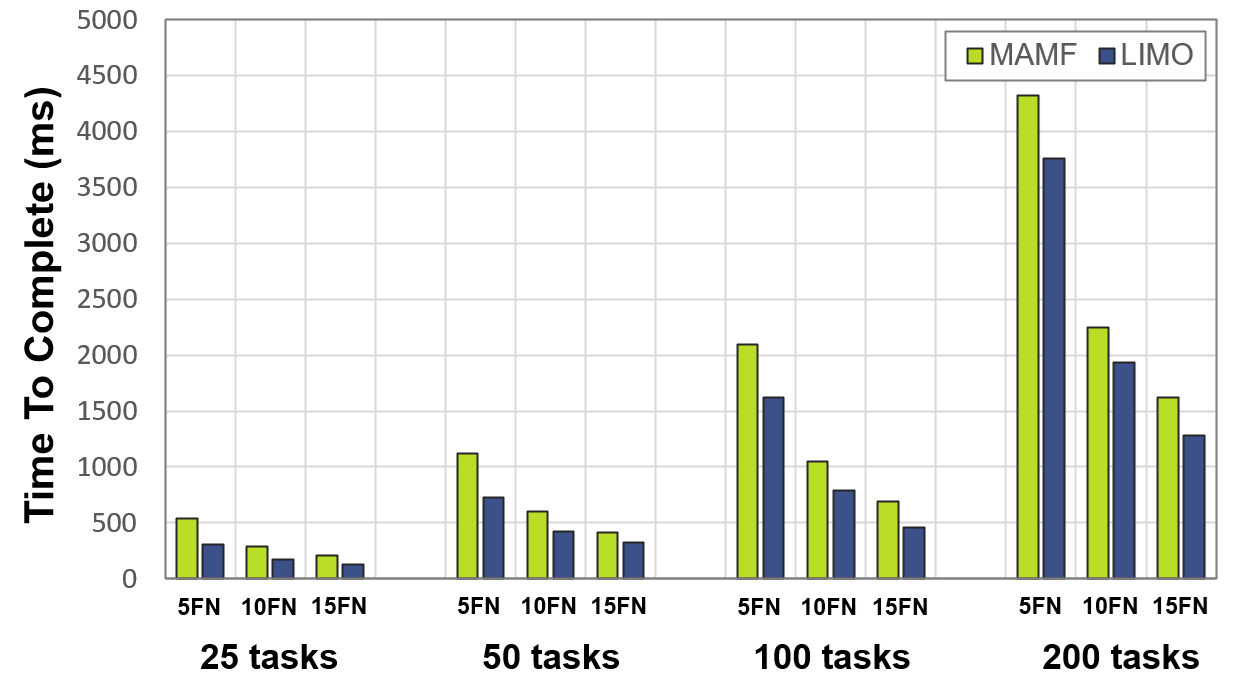}}
\caption{Average TTC of Tasks for different number of Fog nodes.}
\label{3_ttc}
\end{figure}

In addition to analyzing the utilization of every one of the fog devices, it is also important to evaluate the achieved average value in the network. Fig. \ref{1_util} illustrates the average utilization of nodes for various combinations of node counts ($FN$) and tasks ($T$). While LIMO tries to distribute the utilization evenly to prevent overloading and the creation of queues, it also reduces the average utilization in the network by 10\% compared with the MAMF offloading mechanism. This outstanding feature could be later used in the design of the fog infrastructures to get the same performance with fewer fog devices in the network.

Given that some fog nodes become overloaded during peak hours in urban areas, it is necessary to analyze the impact of increasing task numbers on the offloading rate, utilization, and TTC. As the number of tasks rises, fog nodes in densely populated urban regions tend to become overloaded. Implementing a mobility-aware load balancing (as in LIMO) prevents this overloading, thereby reducing the number of tasks required to be migrated to the cloud. When the number of tasks is low, overloading does not occur, making it unnecessary to measure the cloud migration. On the other hand, as the task numbers increase, we need to evaluate the improvements made by the LIMO against the MAMF offloading mechanism, in terms of reduction in the cloud migration. Fig. \ref{2_OL} illustrates the average offloading rate of tasks to both fog nodes and the cloud, while using MAMF, and LIMO offloading techniques. As observed, the LIMO method results in fewer tasks being offloaded to the cloud by 15\%, with a greater number of tasks being processed within the fog layer. This reduction in the cloud offloading rate contributes to a decrease in the amount of TTC. As shown in Fig. \ref{3_ttc}, the TTC in  LIMO is 18\% lower compared to the MAMF method, particularly as the number of tasks increases, a decreasing trend is observed.

\section{Conclusion and Future Studies}
Many QoS parameters, e.g., resource utilization, throughput, cost, response time, performance, and energy consumption can be improved by providing an even balance of load in fog networks. In this study, we first investigated the challenges faced by fog computing environments in providing mobility support, and then proposed LIMO; a mobility-aware autonomous task migration policy based on a combination of autonomous computing principles. LIMO utilizes the MAPE control loop and initiates module migration when the user's distance from the fog node exceeds a prespecified value. By employing the PSO metaheuristic algorithm, LIMO conducts load balancing by exploring the solution space and selecting the best fog node for migration when required. The proposed model ensures the QoS for end users by migrating application modules that operate based on MAPE control loop concepts. This approach can be used for preemptively determining application module migrations, thereby reducing service downtime. The evaluations indicate that LIMO significantly reduces the average execution delay, network usage, and collision incidents.

Several research opportunities could be considered as our future studies. These include enhancing the robustness of the proposed model, in terms of security and privacy by including them in the MAPE loop, consideration of green energy consumption and sustainability, considering the mobility of edge or fog nodes (as opposed to user mobility), and finally, employing deep learning techniques alongside heuristic methods to better optimize the problem.
\bibliographystyle{IEEEtran}
\bibliography{references}
\end{document}